# Low-energy electron attachment to $NO_2$: absolute cross sections

Ana I. Lozano[a,b], Francisco Blanco[c], Juan C. Oller[d], Paulo Limão-Vieira[e], and Gustavo García[a,*]

[a] *Instituto de Física Fundamental, Consejo Superior de Investigaciones Científicas, Serrano 113-bis, 28006 Madrid, Spain*

[b] *Institute de Recherche en Astrophysique et Planétologie (IRAP), Université Toulouse III - Paul Sabatier, 9 Avenue du Colonel Roche, Toulouse 31028, France*

[c] *Departamento de Estructura de la Materia, Física Térmica y Electrónica eIPARCOS, Universidad Complutense de Madrid, 28040 Madrid, Spain*

[d] *División de Tecnología e Investigación Científica, Centro de Investigaciones Energéticas, Medioambientales y Tecnológicas, 28040 Madrid, Spain*

[e] *Laboratório de Colisões Atómicas e Moleculares, CEFITEC, Departamento de Física, Faculdade de Ciencias e Tecnologia, Universidade NOVA de Lisboa, Caparica 2829-516, Portugal*

ABSTRACT.

Resonances from electron attachment to $NO_2$ have been identified in the total electron scattering cross sections (TCS) measured with a magnetically confined electron transmission apparatus. The corresponding absolute electron attachment cross sections have been derived from the TCS values through a self-consistent scattering channel deconvolution process. The positions of some of these resonances are consistent with previous elastic scattering calculations and dissociative electron attachment experiments. However, both the observed positions and the magnitude of the present resonance cross sections are in contradiction with the available recommended TCS values thus suggesting that electron scattering cross section databases need to be updated.

## 1. Introduction

Due to the scientific and technological relevance of nitrogen oxides in nature, electron scattering cross sections from $N_xO_y$ (x,y=1,2) have been extensively studied from both the experimental and theoretical points of view. Results of these studies up to 2001 were compiled by Karwasz et al. [1] and later complemented by Song et al. [2] in 2019. In particular, $NO_2$ plays an important role in atmospheric chemistry as one of the most active air pollutants [3], causing health issues [4]. In recent years, an important effort has been made to identify possible radiosensitizers containing the $NO_2$ group in their molecular structure [5]. These compounds have demonstrated their capability to reduce tumor hypoxia in radiotherapy treatments via NO generation under low-oxygen conditions. Different strategies to generate nitric oxide during tumor irradiation have been investigated in nanomedicine [6] but probably the most efficient way to generate NO from nitrite compounds, on a molecular level, is the dissociative electron attachment (DEA) process. DEA to $NO_2$ has been studied by different authors by analyzing the anionic fragments resulting from the interaction of low-energy electrons with such molecules. In particular $O^-$ formation yields from low energy electron collisions with $NO_2$ molecules have been reported by different experimental groups [7-13]. These dissociative processes always occur after a resonant temporary anion formation by electron attachment (EA) to the target molecule. Some of these EA resonances were identified in early elastic electron scattering cross section calculations by Curik et al. [14] using a single center expansion (SCE) approach. Through a detailed analysis of their potential curves, they concluded the presence of three electron attachment shape resonances labeled by them as $a1$, $b2$, and $b1$ irreducible representation (IR) states (see Ref. [14] for details). Subsequently, Munjal et al. [15] used the R-matrix method to calculate elastic, electronic excitation and total electron scattering cross sections for $NO_2$ for impact energies from 0 to 12 eV. They found two low-lying shape resonances of $^3B_1$ and $^1B_1$ symmetries along with one Feshbach and eight core-excited resonances of different symmetries at higher energies. Gupta et al. [16], also with the R-matrix method, found similar results, although the main two low energy resonances were slightly shifted to higher energies. More recently, Liu et at. [17] modeled the vibrational dynamics of $NO_2$ and analyzed the attachment along each normal mode, providing the position and width of these two resonances for which EA and DEA cross sections were reported. Despite these theoretical and experimental efforts some inconsistencies in the positions and intensities (magnitudes) of the resonances persist. Additionally, the correlation between the calculated EA cross sections and the observed DEA fragmentation is not clearly addressed in previous publications.

The aforementioned discrepancies can be examined from an alternative perspective. Processes involving electron attachment are expected to manifest as resonances in the total electron scattering cross section (TCS) as a function of the incident electron energy. Accurate TCS measurements can be utilized to determine the location of these resonances and yield their related absolute cross section values. According to the recent analysis by Song et al. [2], the TCS reported by Szmytkowski et al. [18, 19] and Zecca et al. [20] can be considered as the most reliable values available in the literature. Consequently, the TCS values suggested in Ref. [2] for electron impact energy values between 1 and 20 eV are obtained from results in Refs. [18-20]. However, these recommended TCS values do not show any resonance in this energy range which clearly contradicts the previous discussion. This observation, along with other inconsistencies in the recommended elastic cross section values, prompted our recently published research on electron scattering cross sections from $NO_2$ for impact energies between 1 and 1000 eV [21]. There we concluded that for the lower impact energies (1-10 eV) additional investigations are needed to establish the connection between the existing local maxima in the experimental TCSs related to temporary anion formation and their subsequent decomposition processes. In the present study we focus on the low energy (0-20 eV) experimental TCS values of Ref. [21] to determine the contribution of electron attachment events to the overall scattering process. The identified EA resonances are contrasted with low energy electron scattering calculations, and their relation to the anion fragmentation observed in DEA experiments is examined.

## 2. Procedure

TCSs have been measured using a magnetically confined experimental setup as detailed in previous publications [21, 22]. We here only mention those aspects that are relevant to the present discussion (see Ref. [21] and references therein for further details). The effective energy resolution was determined to be around 50 meV and the total uncertainty associated to these measurements is established to be within 5%. As discussed in Ref. [21] these results are also affected by a systematic error due to the electron scattering into the acceptance angle of the detector (missing angles). The major contribution to this effect arises from elastic and rotational excitation processes, which act as a systematic error source and can only be quantified through external validations, usually calculations. The present TCS values are not corrected for this effect, but we used our IAM-SCAR+I method [23, 24] to estimate its magnitude (see Ref. [21] for details). For the low energies considered here, we assumed that the TCSs are formed by the combination of resonant and non-resonant mechanisms. The former exhibits distinct peak features at certain energies and has been attributed to electron attachment events, whereas the latter demonstrates a continuous variation with energy and is primarily linked to elastic and electronic excitation processes alongside ionizing collisions above 9.6 eV. Resonances were extracted from the TCSs by removing the continuum-background component. More details on this subtraction procedure are given in the next section.

**3. Results and discussion**

The current experimental electron scattering TCSs for impact energies ranging from 1 to 20 eV are shown in Table 1. This table also presents the experimental missing angles and the corresponding TCS corrections derived from the IAM-SCAR+I [24] calculated differential elastic and rotational excitation cross sections. While this method is not suitable for such low energies, where it tends to overestimate the cross-section values, it offers an upper limit for the previously mentioned systematic error.

**Table 1.** Present experimental TCS results (±5%) in units of $10^{-20}$ m$^2$ along with the missing angles ($\Delta\theta$) in degrees and the estimated contribution of the elastic and rotational scattering into the missing angles to the TCS (in units of $10^{-20}$ m$^2$) as calculated with our IAM-SCAR+I method, $\sigma(\Delta\theta)$, in units of $10^{-20}$ m$^2$.

| **E(ev)** | **$TCS_{exp}$** | **$\Delta\theta$** | **$\sigma(\Delta\theta)$** | **E(eV)** | **$TCS_{exp}$** | **$(\Delta\theta)$** | **$\sigma(\Delta\theta)$** |
|---|---|---|---|---|---|---|---|
| 1.0 | 13.6 | 22.8 | 3.4 | 6.6 | 14.0 | 8.7 | |
| 1.2 | 14.3 | 20.7 | | 6.8 | 13.5 | 8.5 | |
| 1.4 | 14.4 | 19.1 | 2.5 | 7.0 | 13.4 | 8.4 | 0.6 |
| 1.6 | 13.5 | 17.8 | | 7.1 | 13.8 | 8.3 | |
| 1.8 | 13.1 | 16.8 | | 7.2 | 14.5 | 8.3 | |
| 2.0 | 13.3 | 15.9 | 1.9 | 7.5 | 14.3 | 8.1 | |
| 2.2 | 14.2 | 15.1 | | 7.8 | 14.5 | 8.0 | |
| 2.4 | 15.0 | 14.5 | | 6.6 | 14.0 | 8.7 | |
| 2.6 | 15.6 | 13.9 | | 8.0 | 14.4 | 7.9 | |
| 2.8 | 16.2 | 13.4 | | 8.2 | 14.7 | 7.8 | |
| 3.0 | 15.6 | 12.9 | 1.6 | 8.5 | 14.9 | 7.6 | |
| 3.2 | 15.2 | 12.5 | | 8.8 | 15.4 | 7.5 | |
| 3.4 | 14.8 | 12.1 | | 9.0 | 16.0 | 7.4 | |
| 3.6 | 14.8 | 11.8 | | 9.2 | 16.2 | 7.3 | |
| 3.8 | 13.3 | 11.5 | | 9.5 | 16.7 | 7.2 | |
| 4.0 | 12.5 | 11.2 | 1.1 | 9.8 | 17.0 | 7.1 | |
| 4.2 | 12.3 | 10.9 | | 10.0 | 16.7 | 7.0 | 0.5 |
| 4.4 | 12.7 | 10.6 | | 10.2 | 16.2 | 6.9 | |
| 4.6 | 12.9 | 10.4 | | 10.5 | 16.0 | 6.9 | |
| 4.8 | 13.1 | 10.2 | | 11.0 | 16.2 | 6.7 | |
| 5.0 | 13.3 | 10.0 | 1.0 | 12.0 | 16.2 | 6.4 | |
| 5.2 | 13.7 | 9.8 | | 13.0 | 16.5 | 6.2 | |
| 5.4 | 13.8 | 9.6 | | 14.0 | 16.7 | 5.9 | |
| 5.6 | 13.5 | 9.4 | | 15.0 | 16.6 | 5.7 | |
| 5.8 | 13.1 | 9.2 | | 16.0 | 16.5 | 5.6 | |
| 6.0 | 13.0 | 9.1 | | 18.0 | 16.5 | 5.2 | |
| 6.2 | 13.2 | 8.9 | | 20.0 | 16.4 | 5.0 | 0.3 |
| 6.4 | 13.6 | 8.8 | | 11.0 | 16.2 | 6.7 | |

Our TCS results are also shown in Figure 1 along with the experimental values from Szmytkowski et al.[18], Szmytkowski and Mojezko [19] and the recommended values from Song et al. [2]

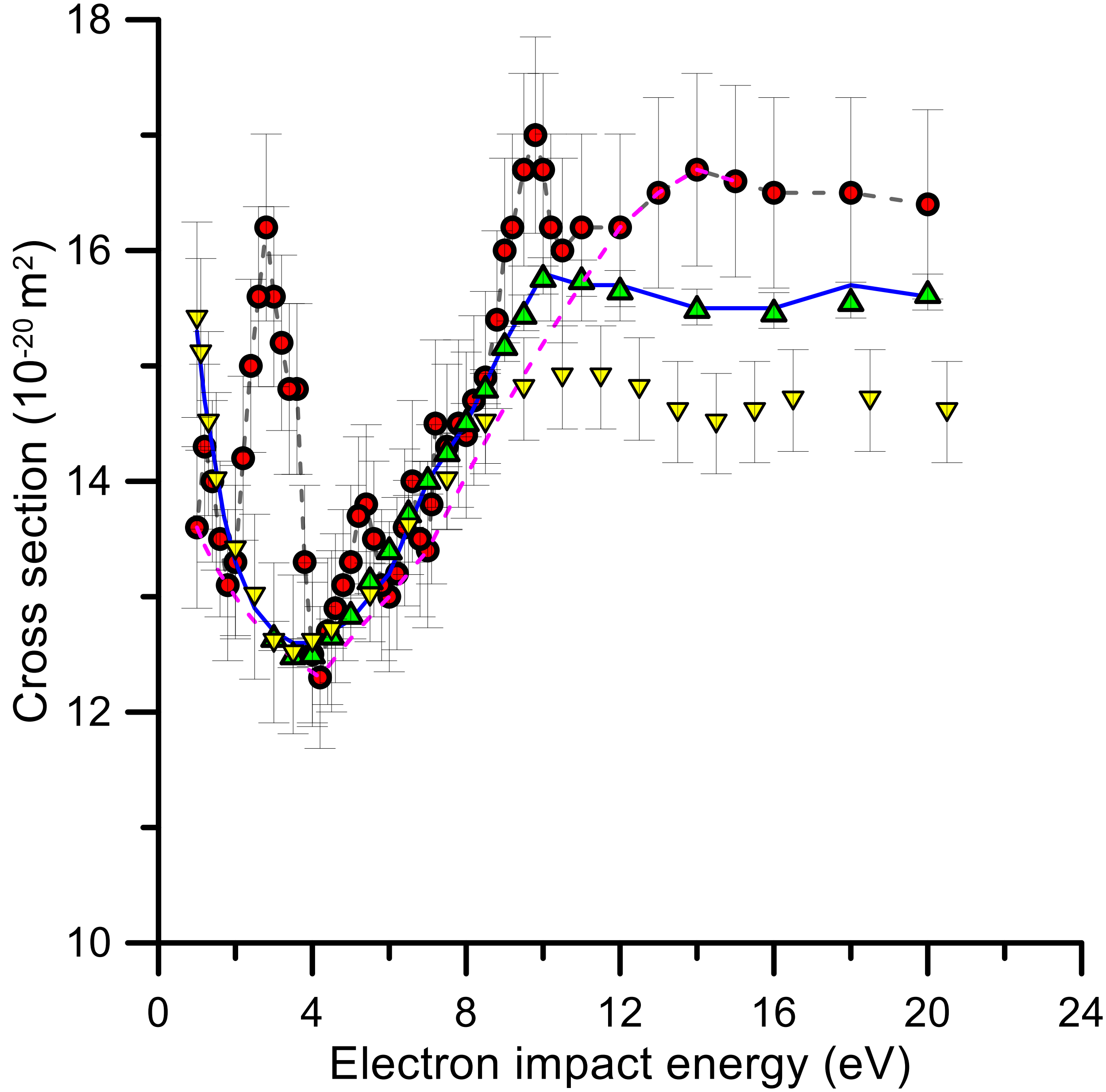

**Figure 1.** Experimental TCS for $NO_2$.--●-- , present results; ▼, experimental values from Ref. [18];

△, Experimental values from Ref. [19]; —, recommended data from Ref. [2]; ---, background cross section (see text for details). Error bars only account for random uncertainties, and all experimental values are not corrected for systematic errors resulting from electron scattering in the forward direction (missing angles).

As can be seen in Figure 1, there are some discrepancies between the present results and those of Refs. [18,19], upon which the recommended values of Song et al.[2] are based. To directly compare these experimental results, the error bars represent only random uncertainties, and all experimental values remain uncorrected for systematic errors caused by electron scattering in the forward direction (missing angles). The local maxima of the TCS that we found between 1 and 10 eV are somewhat absent in the experimental results of Refs. [18,19]. Furthermore, above 10 eV our TCS values tend to be higher in magnitude than theirs, but discrepancies are within the combined uncertainty limits. The absence of resonances in previous measurements, and consequently in the recommended data of Ref. [2], can be attributed to a lack of energy resolution. The energy spread of the electron beam used in Refs. [18, 19] is quoted to be 80 meV. However, the electrons emerging from the gas cell are energy analyzed using a retarding field element (RPA), which defines the effective energy resolution of the entire system. Unfortunately, this is not discussed in Refs. [18, 19], and based on the present results, we assume that the actual energy resolution in those measurements is likely worse than 80 meV. A key benefit of employing a magnetically confined electron beam system like ours is that the electron beam maintains zero divergence and impacts the RPA barrier at a right angle, whereas the divergence angle of a free electron beam decreases the effectiveness of these energy analyzers.

To analyze the position and structure of all resonances observed in our TCS measurements, we have subtracted the continuum background cross-section —mainly due to elastic and electronic excitation processes— from the total cross section for impact energies ranging from 1 to 15 eV.

To accomplish this, we regarded the TCS values on either side of each resonance or group of resonances as the background cross sections. The gaps between these points were smoothly interpolated in a log-log graph.

The results obtained are plotted in Figure 2 and the resultant cross sections are assigned to electron attachment (EA) processes. A standard Gaussian fit with a commercial software (MagicPlotStudent) of the ensuing structures allowed the identification of different resonances. The uncertainty in the absolute value of these cross sections has been estimated to be about ± 15% by adding in quadrature the uncertainty limits of the cross sections involved in the subtraction procedure. If accounting for the missing angle correction, this 15% would rise to 25% for the lowest energy resonance. The uncertainty in the resonance positions is established as ± 0.1 eV due to the combination of energy resolution and the energy difference between the measured energy points.

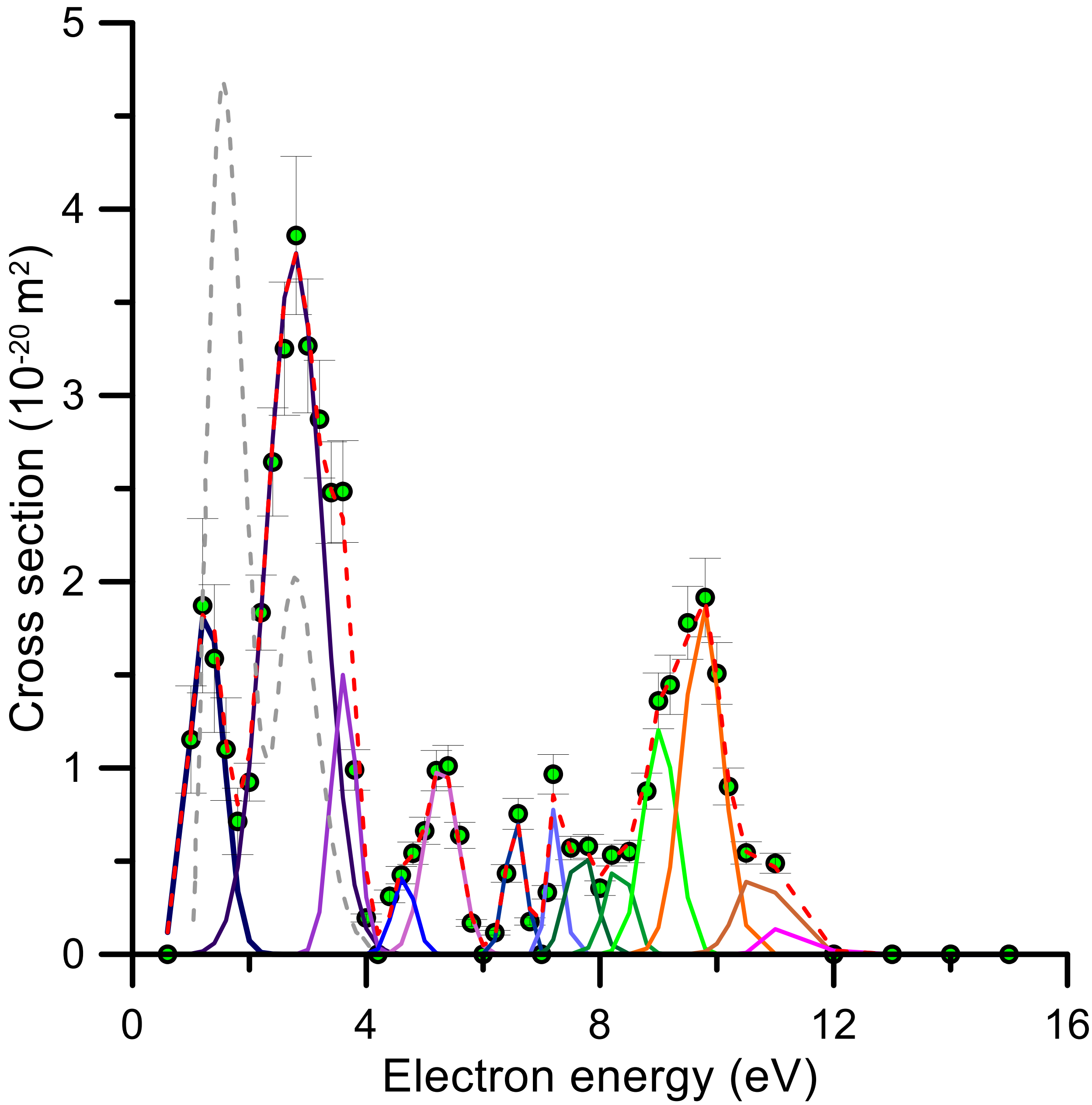

**Figure 2.** Electron attachment cross sections: ●, present results (see text for details); ---, Gaussian fit to the present experimental data. ---, calculations from Liu et al. [17]

The peak position and the corresponding EA cross section assigned in this study together with previous data available in the literature are shown in Table 2.

To our knowledge, no previous direct measurements of EA resonance positions and shapes have been reported in the literature. However, as shown in Table 2, several calculations have been performed using different approaches. Curik et al. [14], using a model potential method, found three resonances due to electron attachment to $NO_2$. Two of them are very broad (2-3 eV width) and are placed at relatively high energies (13.7 and 18.1 eV, respectively) where electronic excitation and ionization channels are relevant and therefore not distinguishable in our total cross section measurements. The third one is located at 2.21 eV with a width of 0.4 eV and could correspond to the intense resonance we found around 2.8 eV. It has been characterized as a $b_1$ IR resonant state although it was not possible to locate it in the corresponding integral cross section calculation (see Ref. [14] for details). More elaborated calculations were performed by Munjal et al. [15] using the R-matrix method and including 21 states. They found two shape resonances of $^3B_1$ and $^1B_1$ symmetries located at 1.18 and 2.33 eV, respectively. These positions agree reasonably well with those of our first two resonances and were calculated in Ref. [15] by assuming that one N-O bond is stretched while the other bond and the bonding angle are frozen. As shown in Table 2, the locations of other core excited resonances described in Ref. [15] agree with some of the present electron attachment features at 5.2, 6.6, 8.2, 9.0, 9.8 and 10.5 eV. Gupta et al. [16], using also the R-matrix procedure, found resonant structures around 1.33, 3.0, 6.7 and 8.41 eV which are also in reasonable agreement with some of the present resonances. The latest calculation by Liu et al. [17], using a relatively simple approach, which combines a normal mode representation of the $NO_2$ vibrational dynamics with a one-dimensional electron attachment for each vibrational mode, reported on the position and width of the two lowest energy resonances at 1.65 and 2.93 eV. These values are consistent with earlier calculations and match our experimental results. Despite the spread of the discussed theoretical results on the position of this prominent lowest energy resonances, the average values are 1.4 ±0.2 and 2.6±0.4 eV, respectively, which agree well with our corresponding values of 1.2 ±0.1 and 2.8±0.1 eV.

Table 2 also presents the positions of the resonances obtained from DEA experiments by measuring the $O^-$ formation yield. It is important to note that once the parent anion is formed, the DEA resulting in $O^-$ formation is one possible decay route but not the only one; autodetachment and other dissociative processes may also occur, and the dynamics of these processes could influence the location of the maximum cross-section. Nonetheless, as observed in Table 2, the $O^-$ formation resonance positions reported by all prior experiments [7-11] are in good agreement with each other and are consistent with the current EA resonances. Rangwala et al. [11] identified three resonances leading to the formation of $O^-$, with their respective peaks at 1.4, 3.1 and 8.3 eV in concordance with our EA resonances. The total negative ion formation signal as a function of electron energy for $NO_2$ reported by Rallis et al. [9] (see Fig. 3 of Ref. [9]) qualitatively agrees with the present peak positions, showing an absolute maximum around 2.8-3.0 eV and two additional local maxima at about 5.6 and 8.1 eV, respectively (see Table 2). As shown in Table 2, some of the resonances reported in Refs. [8] and [10] correspond to $NO^-$ and $O_2^-$ formation, although their relative peak intensities were found to be at least two orders of magnitude lower than those corresponding to $O^-$ (see Ref. [8]). Note that the peak we observed at around 4.6 eV which is supposed to be related to $NO^-/O_2^-$ formation is of the same order of magnitude as those related to $O^-$ formation. This may indicate that autodetachment and neutral dissociation processes are strongly competing with these anion dissociation channels. For $CO_2$, Dvořák et al. [17] described certain fine structures seen in the electron energy loss spectra as a vibrational pseudocontinuum arising from vibronic coupling between the $^2\Pi_u$ resonant state of the $CO_2^-$ anion and the vibrational structure of the $^2\Sigma_g^+$ of the $CO_2$ molecule, facilitated by the transformation of the incoming resonant *p* wave to the outgoing *s* wave. There is no evidence of a similar process for $NO_2$, and the complex model they used is so fitted to the structure of $CO_2$ that we consider this result cannot be extrapolated to the present case. In any case, this fact does not contradict the previous statement; it simply strengthens the idea that autodetachment processes are prevalent in these resonances.

As mentioned earlier, the current absolute EA cross sections for $NO_2$ are shown in Figure 2. The corresponding numerical data are also presented in Table 2. No previous experimental EA cross sections have been reported in the literature. Liu et al. [17], using the aforementioned approach, calculated the absolute EA cross sections for the two lowest energy resonances. Their results are displayed in Table 2 and Figure 2 for comparison. Although we agree on the order of magnitude of the cross sections, their maximum value for the first resonance is over twice ours and that for the second resonance is half of ours. Currently, we are unable to explain this inconsistency. The initial resonance at 1.2 eV is indeed close to the minimum energy operating limit of our experimental setup [21], and its cross section may be slightly undervalued; however, for the subsequent one at 2.8 eV, no further systematic errors should be expected. Elucidating this aspect would necessitate additional theoretical and experimental studies.

Concerning DEA experiments, only Rangwala et al. [11] reported absolute cross sections for $O^-$ production. These results were derived after a time-of-flight (TOF) analysis of the anion species generated by the interaction of a magnetically collimated pulsed electron beam with an effusive $NO_2$ molecular beam. The absolute values were evaluated through a relative flow technique. This is an indirect method; therefore, in addition to the 13% statistical uncertainty quoted by the authors, a total error around 20% can be expected for these absolute cross-section values [11]. Regarding the first two resonances, when we compare our EA cross sections with the Rangwala's [11] DEA for $O^-$ production cross sections we find ours are more than one order of magnitude higher than theirs. A similar situation was found by Liu et al. [17], as their calculated EA cross sections were determined to be 46 times larger than the experimental DEA cross sections; thus verifying that the majority of anions formed by electron attachment are eliminated through autodetachment processes. Moreover, Liu et al. [17] modeled a survival function to calculate the fraction of EA generated anions that undergo DEA producing $O^-$.Their modeled DEA cross sections were found 2-10 times larger than the experimental ones.

**Table 2.** Observed electron scattering from $NO_2$ resonances (units eV) including the estimated electron attachment (EA) cross sections (units of $10^{-20}$ $m^2$). The resonance positions (units eV) from available calculations and dissociative electron attachment measurements are also included.

| **Resonance** | **Peak position** (eV) | | | **EA cross section** ($10^{-20}$ $m^2$) | |
|---|---|---|---|---|---|
| | Present work | Other experiments | Other calculations | Present work | Other calculations |
| $O^-/NO_2$ ($^3B_1$ shape) [15, 16, 24] | 1.2 | 1.4 [11], 1.8 [8, 10, 12, 13], 1.9 [7] | 1.18 [15], 1.33 [16], 1.65 [17] | 1.87 | 4.69 [17] |
| $O^-/NO_2$ ($^2B_1$ shape) [16] | 2.8 | 3.0 [7, 9], 3.1[11], 3.5 [12, 13] | 2.21[14], 2.33 [15], 3.0 [16], 2.93 [17] | 3.86 | 2.02 [17] |
| $NO^-/NO_2$ ($^1B_1$ shape) [15] | 3.6 | 3.2 [10], 3.3 [8], | | 2.48 | |
| $O_2^-/NO_2$ | 4.6 | 4.3 [8] | | 0.42 | |
| ($^3B_2$ Core excited) [15] | 5.2 | | 5.1 [15], 5.32 [15] | 0.99 | |
| ($^1A_1$ Feshbach) [15] ($^1A_2$ core excited) [15] | 6.6 | | 5.57 [15], 5.84 [15], 6.7 [16] | 0.75 | |
| | 7.2 | | | 0.97 | |
| | 7.8 | | | 0.58 | |
| $O^-/NO_2$ ($^3B_2$ core excited)[15] | 8.2 | 8.1 [9], 8.3 [11], 8.5 [12, 13] | 8.3 [15], 8.41 [16] | 0.53 | |
| ($^3A_1$ core excited)[15] | 9.0 | 8.75 [7] | 9.1 [15] | 1.36 | |
| ($^3A_2$ core excited)[15] | 9.8 | | 9.8 [15] | 1.91 | |
| ($^1A_2$ core excited)[15] | 10.5 | | 10.2 [15] | 0.54 | |
| | 11.0 | | | 0.49 | |

**4. Conclusions**

In summary, we have reported new experimental results on the total electron scattering cross sections of $NO_2$ for impact energies ranging from 1 to 20 eV, as measured with a magnetically confined electron-beam-transmission apparatus. The reduced uncertainty limits (±5%) and effective energy resolution (50 meV) of the present experimental conditions allowed the extraction of the resonant EA cross section contribution to the TCS. Some of these resonances have been characterized by comparing their respective positions with those predicted by previous R-matrix calculations. This finding resolves the inconsistency between the recommended TCS values of Ref.[2], where no resonances were observed, and the resonant features predicted by these calculations [14, 15, 17]. In addition, most of the present EA resonances are confirmed by previous DEA experiments [7-13]. Our absolute EA cross section values exceed by over one order of magnitude the respective DEA cross sections reported by Rangwala et al. [11]. Similar differences have been observed in recent EA cross section calculations [17], thereby confirming that subsequent autodetachment processes prevail in the electron attachment dynamics of $NO_2^-$.

As a consequence of the present study, it has become evident that further theoretical calculations and experimental investigations are required to achieve a more comprehensive characterization of the temporary anionic states of $NO_2$ and their respective decay pathways, whether via dissociative electron attachment (DEA) or autodetachment, and the role of the latter in subsequent neutral fragmentation processes. Moreover, the data currently recommended in Ref. [2] should be revised and updated to include the current EA resonances and cross section values along with recently reported data on electron scattering cross sections [21].

**CRediT authorship contribution statement**

The manuscript was written through contributions of all authors. All authors have given approval to the final version of the manuscript. All authors contributed equally.

**Declaration of competing interest**

The authors declare that they have no known competing financial interests or personal relationships that could have appeared to influence the work reported in this paper.

**Acknowledgments**

This study has been performed within the framework of the 21GRD02 BIOSPHERE project supported by the European Association of National Metrology Institutes (EURAMET). It has also been funded by the Spanish Ministerio de Ciencia e Innovación, project PID2019-104727RB-C21. P.L.-V. acknowledges the Portuguese National Funding Agency (FCT) research grant CEFITEC (UIDB/00068/2020).